\documentclass[a4paper,12pt]{article}
 \usepackage{amssymb,amsmath}
\usepackage{amsmath,amsfonts,amssymb,amsthm,mathtools}
\usepackage{tikz} 
\usetikzlibrary{arrows, backgrounds, patterns,matrix,shapes,fit,calc,3d,shadows,plotmarks}
\usepackage{tikz-network}

\usepackage{tabu}
\usepackage{caption}
\usepackage{url}
\usepackage[colorlinks=true,urlcolor=blue,linkcolor=black,citecolor=black]{hyperref}
\usepackage{booktabs}
\usepackage{algorithm}
\usepackage{algpseudocode}

\DeclareCaptionType{equ}[][]

\newcommand{\beqn}[1]{\begin{eqnarray}\label{#1}}
\newcommand{\eeqn}{\end{eqnarray}}
\newcommand{\beq}{\begin{eqnarray*}}
\newcommand{\eeq}{\end{eqnarray*}}
\newcommand{\bit}{\begin{itemize}}
\newcommand{\eit}{\end{itemize}}

\newcommand{\comment}[1]{}

\title{Forward-Backward Binarization}

\author{
  Ismail Belgacem\thanks{Corresponding author: \href{mailto:ismail.belgacem.81@gmail.com}{ismail.belgacem.81@gmail.com}} \and
  Franck Delaplace
}

\date{
  IBISC Lab, Paris-Saclay University, Évry\\
  IBGBI 23, boulevard de France, 91037 Évry, France
}

\begin{document}

\maketitle


\begin{abstract}
	Binarization of gene expression data is a \textbf{critical prerequisite} for the synthesis of Boolean gene regulatory network (GRN) models from omics datasets. Because Boolean networks encode gene activity as binary variables, the accuracy of binarization directly conditions whether the inferred models can faithfully reproduce biological experiments, capture regulatory dynamics, and support downstream analyses such as controllability and therapeutic strategy design. In practice, binarization is most often performed using thresholding methods that partition expression values into two discrete levels, representing the absence or presence of gene expression. However, such approaches oversimplify the underlying biology: gene-specific functional roles, measurement uncertainty, and the scarcity of time-resolved experimental data render thresholding alone insufficient. To overcome these limitations, we propose a novel \textbf{regulation-based binarization method} tailored to snapshot data. Our approach combines thresholding with functional binary value completion guided by the regulatory graph, propagating values between regulators and targets according to Boolean regulation rules. This strategy enables the inference of missing or uncertain values and ensures that binarization remains biologically consistent with both regulatory interactions and Boolean modeling principles of the gene regulation. Validation against ODE simulations of artificial and established Boolean GRNs demonstrates that the method achieves accurate and robust binarization, thereby  strengthening the reliability of Boolean network synthesis.
\end{abstract}

\section*{Introduction}

\begin{figure}[h]
	\centering
	\includegraphics[width = 0.95\textwidth]{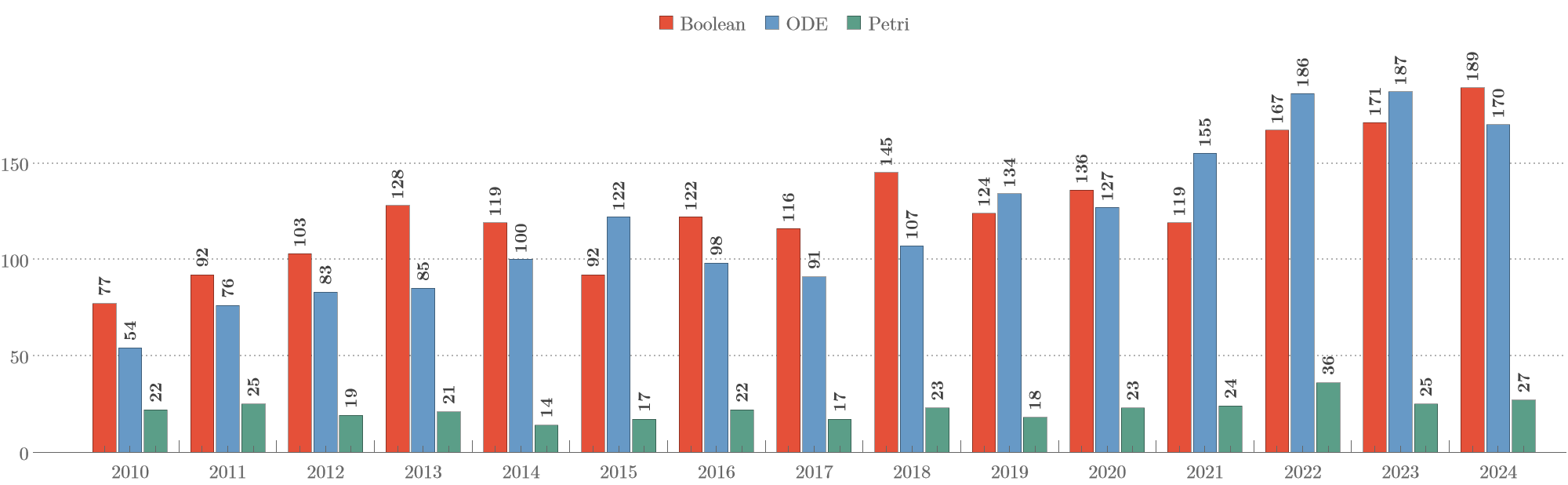}
	
	\medskip
\begin{minipage}{0.9\textwidth} \small
	The queries are applied on title and abstract and they are respectively:
	\texttt{Boolean AND (Model OR Network)} for Boolean, \texttt{ (ODE OR ordinary differential equation) AND Model} for ODE, and \texttt{Petri AND Net} for Petri net.
\end{minipage}
\caption{Number of publications since $2010$ on \textsc{pubmed} related to modeling frameworks.} 
\label{fig:model-publication}
\end{figure}

Boolean network represents a discrete modeling framework for gene regulatory networks (GRN) that can be regarded as a gold standard in the field of biological modeling, as evidenced by the number of published studies adopting this approach compared to alternative modeling paradigms (Figure~\ref{fig:model-publication}). Boolean network modeling  have proven useful in providing relevant biological insights and discovering therapeutic strategies and drugs. Drug design is generally based on the controllability, namely the ability of automatically inferring which targets should be blocked to purposely  deviate the cell fate \cite{pardo2021, biane2019, angarica2017bioinformatics, zickenrott2016prediction}

In the literature, Boolean network synthesis typically relies on binary gene expression profiles as input~\cite{shi2020aten, aghamiri2021taboon, chevalier2024bonesis}.

However, gene expression data are inherently continuous rather than binary. In practice, gene expression experiments quantify either mRNA levels or the abundance of gene products, most often proteins. For instance, northern blot assays and quantitative nuclease protection assays (qNPAs) measure mRNA levels in biological samples~\cite{streit2009northern, bourzac2011high}. Protein abundance can be assessed using several techniques, including protein microarrays~\cite{lueking1999protein}, western blotting~\cite{sinkala2017profiling, mckiernan2008protein, pillai2020systematic}, enzyme-linked immunosorbent assays (ELISAs)~\cite{joos2000microarray}, and reverse-phase protein arrays~\cite{charboneau2002utility, mannsperger2010rnai, akbani2014realizing}.

More broadly, high-throughput biochemical assays enable multi-omics analyses by simultaneously quantifying molecules from the genome, transcriptome, proteome, metabolome, and epigenome~\cite{conesa2019making}. A wide range of experimental technologies has been developed for measuring gene expression, including RNA-seq, gene expression microarrays, fluorescence flow cytometry, and fluorescence microscopy~\cite{arikawa2008cross, hilario2007protocols, jin2012rna}. Given this diversity of methods, providing an exhaustive overview is challenging, as each technology differs in its principles, resolution, and data output.

In the literature, Boolean network synthesis typically relies on binary gene expression profiles as input~\cite{shi2020aten, aghamiri2021taboon, chevalier2024bonesis}. 

However, gene expression data are inherently continuous rather than binary. Gene expression experiments generally quantify either mRNA levels or the abundance of gene products, most often proteins. For example, northern blot assays and quantitative nuclease protection assays (qNPAs) measure mRNA levels in biological samples~\cite{streit2009northern, bourzac2011high}, while protein abundance can be assessed using techniques such as protein microarrays~\cite{lueking1999protein}, western blotting~\cite{sinkala2017profiling, mckiernan2008protein, pillai2020systematic}, enzyme-linked immunosorbent assays (ELISAs)~\cite{joos2000microarray}, and reverse-phase protein arrays~\cite{charboneau2002utility, mannsperger2010rnai, akbani2014realizing}. Moreover, high-throughput biochemical assays enable multi-omics analyses that simultaneously measure molecules from the genome, transcriptome, proteome, metabolome, and epigenome~\cite{conesa2019making}. A wide range of technologies has been developed to measure gene expression, including RNA-seq, gene expression microarrays, fluorescence flow cytometry, and fluorescence microscopy~\cite{arikawa2008cross, hilario2007protocols, jin2012rna}. Given this diversity, providing an exhaustive overview of all measurement methods is challenging.

Besides, gene expression datasets used for Boolean network inference can take the form of either time-series data or instantaneous measurements (snapshots), as illustrated by temporal and snapshot RNA-seq profiling in~\cite{chu2016single}. Time-series data record gene expression dynamics over time, with measurements collected at regular intervals (minutes, hours, days), and values typically reported as real numbers or integer counts representing mRNA abundance~\cite{arbeitman2002gene, cho1998genome, bachmann2011division}. In contrast, snapshot data capture gene expression at a single time point under specific conditions, such as healthy versus diseased states, as in the qRT-PCR snapshot dataset described in~\cite{kouno2013temporal}.

Consequently, when synthesizing Boolean networks, a crucial preprocessing step consists in converting continuous expression data into binary values. This process, known as \emph{binarization}, is essential to represent gene regulatory states (0 or 1) and to infer Boolean networks from the resulting binary data. Choosing an appropriate binarization method is therefore critical to ensure that the inferred Boolean models faithfully reflect the underlying biological processes.

Several gene expression binarization methods have been proposed \cite{tuna2010inference, beal2019personalization, jung2017refbool}. Binarization transforms gene expression measurements into indications of whether a gene is active ($1$) or inactive $0$. The most common approach is based on the identification of a threshold delineating the frontier of gene activity: below the threshold the gene is considered as $0$ (inactive) while above it is $1$ (active). However, such approach 
does not account to  the  genes' functional roles behind the Boolean value.Indeed, the key distinction between real-valued data and binary data is that $0$ and $1$ are meant to represent the regulatory activity of a gene.  When a gene is assigned $1$, its regulatory function is considered active, allowing it to influence its downstream targets. 
Conversely, a value of $0$ implies that the gene is unable to exert any regulatory effect on its targets. 
This notion of functional activity introduces an inherent zone of uncertainty that cannot be fully captured by a single expression threshold.

Moreover, existing methods often assume ideal gene expression datasets, such as dense time-series with fine-grained measurements. 
In practice, experimental data are frequently noisy and sparse, often consisting of only a few snapshots. 
To address these limitations, we propose an original and improved approach: a novel method that explicitly incorporates gene regulatory relationships. 
Our method is designed to work with instantaneous data, even from a single steady-state snapshot. 
It combines thresholds estimated from gene expression data with functional binary state completion guided by the regulatory graph. 
During traversal of this graph, binary states are propagated from regulators to their targets and reciprocally, according to Boolean regulation rules.

The paper is organized as follows: we first present a classification of existing binarization approaches and analyze their performance~(Section~\ref{sec:state of art}). We then describe our proposed binarization method and detail the main steps of its algorithm (Section~\ref{sec:Bi4Back}). Next, we evaluate the algorithm on real gene expression data to demonstrate its effectiveness by verifying the correctness of gene binarization using \textsc{ode} simulations of artificial gene regulatory networks or well-known Boolean biological networks. We conclude with a discussion on suitable datasets for reliable binarization and on the potential applications of our method for disease treatment.

\section{State of the Art}
\label{sec:state of art}
Binarization converts continuous gene expression datasets into Boolean values $0$ and $1$.  The predominant methodology in the literature employs a \textbf{threshold-based approach}, where a cutoff value serves as the decision boundary: gene expression levels falling below this threshold are classified as inactive (assigned value $0$), while those exceeding it are considered active (assigned value $1$).

Genes exhibit varying levels of expression within cells, with some producing very few transcripts and thus requiring low detection thresholds to be accurately identified, while others are only expressed above a certain abundance level and remain undetectable below this threshold. Once synthesized, gene products such as proteins or RNAs must identify their specific targets, migrate to DNA binding sites, and bind to regulate transcription \cite{widder2007dynamic, rastogi2018accurate, le2018comprehensive}.

The effectiveness of gene regulation fundamentally depends on the \textbf{binding affinity} -- the strength of molecular interaction between transcription factor proteins and their corresponding DNA target sequences -- a parameter that varies significantly across different genes \cite{widder2007dynamic}. This gene-specific variation necessitates individualized threshold determination, making \textbf{gene-by-gene quantization} approaches more biologically meaningful than uniform global thresholding strategies. Furthermore, the regulatory transition between active and inactive states introduces additional complexity, as the switching mechanism near threshold boundaries involves inherent uncertainty -- the conversion between regulatory states ($0$ and $1$) occurs gradually rather than instantaneously, depending on the underlying molecular binding dynamics \cite{rastogi2018accurate}. To address cases where gene expression levels fall within ambiguous ranges that do not clearly correspond to either active or inactive states, researchers have introduced an \textbf{intermediate classification (NA)} that explicitly accounts for this biological uncertainty \cite{le2018comprehensive}.

Binarization methods incorporate this complexity through \textbf{dual-threshold} methods that partition gene expression data into three distinct categories: inactive ($0$), active ($1$), or unassigned (NA) states.  The performance of such binarization methods relies on algorithms that accurately compute these thresholds. These approaches can be broadly categorized into different classes: single threshold identification, clustering methods, density distribution estimation methods, and fast dynamics detection methods.

\subsection{Single threshold identification}
In the study by Becquet et al. \cite{becquet2002strong}, binarization is  performed using two main approaches. The first approach employs a \textbf{mid-range threshold method}, where the mid-range value (the middle point of the data range) serves as a decision threshold. The second approach uses a \textbf{percentage-based threshold method}, which relies on a predefined percentage of the highest values. For example, the top $30\%$  of values are selected by first ranking all values in ascending order, then setting values that fall within the top 30\%  to $1$, while all other values (the bottom $70\%$) are set to $0$. The threshold denotes a cutoff point used to make binary decisions, and percentile indicates a value below which a certain percentage of data falls. In this approach, the intermediate state is not accounted. 

\subsection{Clustering Methods}
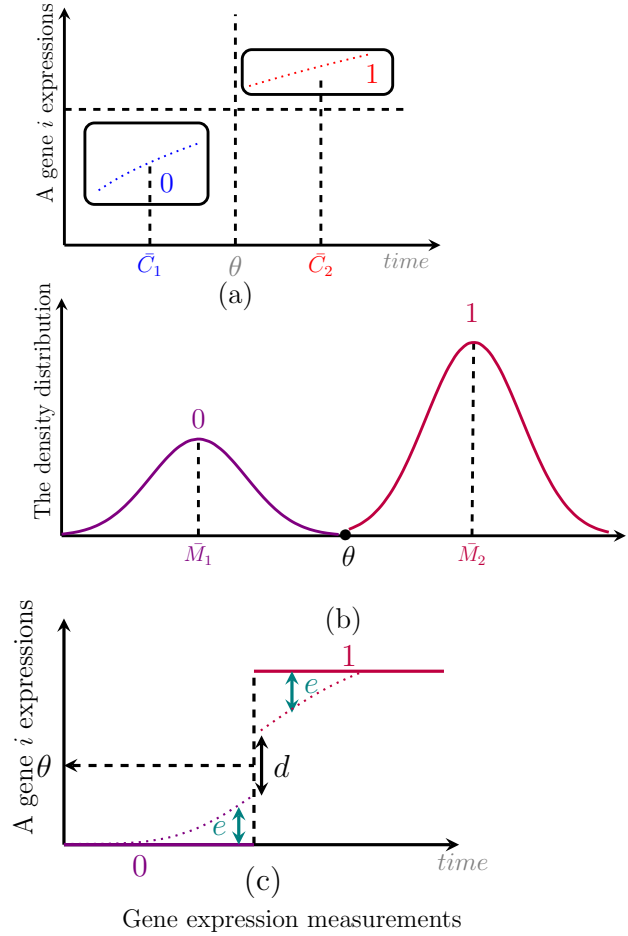
\begin{figure}[htb!]%
\begin{center}
   \begin{tikzpicture} 
   \node at (-2,0) [scale=0.9] {
     \begin{tikzpicture} 
   \node at (5,0) [name=1, color=gray , scale=0.8, below] {$time$} ; 
   \node at (-0.25,0.5) [name=2, scale=0.8, above]{\rotatebox{90}{A gene $i$ expressions}} ;
    \node at (1.5,1.2) [name=1, color= blue, scale=1, below] {$0$} ; 
   \node at (4.5,2.25) [name=2, color= red, scale=1, above]{$1$} ;
   \draw[->, very thick, >=stealth] (0,0)-- (5.5,0);  
   \draw[->, very thick, >=stealth] (0,0)-- (0,3.5);  
	 \draw[very thick,dashed] (0,2) -- (5,2) ;
	 \draw[ very thick, dashed] (2.5,0) node[gray, below] {$\theta$}  -- (2.5,3.4) ;
   \draw[ very thick, dashed] (3.75,0)  node[red, below, scale=0.8] {$\bar{C_2}$} -- (3.75,2.5) ;
    \draw[ very thick, dashed] (1.25,0)  node[blue, below, scale=0.8] {$\bar{C_1}$} -- (1.25,1.25) ;
	  \draw[color=blue, thick, dotted]  plot[domain=0.5:2,
        samples = 8,
        smooth](\x,{sqrt(\x)+0.1});	        
        \draw[color= red, thick, dotted]  plot[domain=2.7:4.5,
        samples = 5, smooth](\x,{sqrt(\x)+0.7});	
       \node at ( 3.7,2.55) [name=data, very thick, draw=black, rounded corners,
                           minimum width=4cm, minimum height=1.2cm,
                           text centered, scale=0.55]{} ; 
 \node at ( 1.2,1.2) [name=data, very thick, draw=black, rounded corners, minimum width=3cm, minimum height=2cm, text centered, scale=0.6]{} ; 
 
\node at (2.5,-0.75){(a)};
	  \end{tikzpicture}
	  
 };
 
   \node at (-0.8,-4) [scale=0.9] { 
   
 \begin{tikzpicture} 
   \draw[<-, very thick, >=stealth] (8.25,0)--(0,0);  
   \draw[->, very thick, >=stealth] (0,0)--(0,3.5); 
   \node at (-0.25,0) [name=2, scale=1, above, scale= 0.8]{\rotatebox{90}{ The density distribution}} ;
    \node at (2,2) [name=1, color= violet, scale=1, below] {$0$} ; 
   \node at (6,3) [name=2, color= purple, scale=1, above]{$1$} ;
	  \draw[very thick, dashed] (2,0) node[violet, below, scale= 0.75] {$\bar{M_1}$} -- (2,1.4)  ;
	    \draw[very thick, dashed] (6,0) node[purple, below, scale= 0.75] {$\bar{M_2}$} -- (6.025,2.9)  ;
   \draw[very thick, dashed] (4.2,0) node[below] {$\theta$};
     \draw[very thick] (4.15,0) node {\textbullet};
	  \draw[ color=violet, very thick, smooth]  plot[domain=0:4.05](\x, {5/(2*sqrt(2*3.14*(0.7^2)))*exp(-(\x-2)^2/(2*(0.7^2))});	       
	   \draw[color=purple, very thick, smooth]  plot[domain=4.20:8](\x, {10/(2*sqrt(2*3.14*(0.7^2)))*exp(-(\x-6.025)^2/(2*(0.7^2))});   
	   
	   \node at (4.12,-1.25){(b)};      
	  \end{tikzpicture}

	  };

\node at (-1.8,-8) [scale=1] { 
	       \begin{tikzpicture} 
   \node at (5.25,0) [name=1, color=gray, scale=0.8, below] {\rotatebox{0}{$time$}} ; 
   \node at (-0.5,0) [name=2,  scale=1, above, scale= 0.85]{\rotatebox{90}{A gene $i$ expressions } } ;
    \node at (1,0) [name=1, color= violet, scale=1, below] {$0$} ; 
   \node at (4,2.5) [name=2, color= purple, scale=1, left]{$1$} ;
   \draw[->, very thick, >=stealth] (0,0)-- (5.2,0);  
   \draw[->, very thick, >=stealth] (0,0)-- (0,3);  
	 \draw[<-, very thick, dashed,>=stealth] (0,1.05) node[ left] {$\theta$} -- (2.5,1.05) ;
	  \draw[very thick, dashed] (2.5,0) -- (2.5,2.3) ;
	  \draw[color=violet, very thick] (0,0)-- (2.5,0) ;
	  \draw[color=purple, very thick] (2.5,2.3)-- (5,2.3) ;
	    \draw[<->, very thick, >=stealth]  (2.6,0.65) --(2.6,1.05)  node[ right] {$d$}  -- (2.6,1.45) ;
	      \draw[<->, very thick, >=stealth, teal] (3,1.75)--(3,2.1)  node[ right,teal] {$e$}  -- (3,2.3) ;
	      \draw[<->, very thick, >=stealth, teal] (2.3,0)--(2.3,0.225)  node[ left, teal] {$e$}  -- (2.3,0.5) ;
	  \draw[color=violet, thick, dotted, samples = 8]  plot[domain=0:2.45, smooth](\x,{(2*\x^4)/(\x^4+3^4)+0.01});	        
        \draw[color= purple, thick, dotted]  plot[domain=2.59:3.9,
        samples = 5,   smooth](\x,{((2*\x^4)/(\x^4+3^4))+0.8});	
       \node at (2.62,-0.5){(c)};   
       \node at (3,-0.75) [name=1, scale=1, below, scale= 0.8] {Gene expression measurements} ;
	  \end{tikzpicture}
	    
	    };
	  
	  \end{tikzpicture}
\caption{(a): A clustering binarization of the expressions of a gene $i$. Where, $\bar{C_1}$ is the center of the first cluster, $\bar{C_2}$ is the center of the second cluster, and $\theta$ is the threshold between the two cluster centers. (b): A bimodal distribution of the gene expression measurements (on the right). Where, $\bar{M_1}$ is the mean of the first distribution, $\bar{M_2}$ is the mean of the second distribution, and $\theta$ is the threshold between the two distribution means. (c): Gene expression measurements over time. Here, $d$ is the distance between two successive measurement points, and $e$ is the error between the measurement point values and a possible step function.}
\label{categories}%
\end{center}
\end{figure}

Binarization can be framed as a \textbf{clustering problem}, where the objective is to partition high-dimensional data into groups with strong internal similarity. For binarization purposes, clustering is implemented as a \textbf{bipartition method}, dividing the data into two distinct clusters respectively corresponding to $0$ and $1$. Using the \emph{k-means algorithm} for gene-by-gene binarization, gene expression measurements over time are separated into two clusters, each characterized by its centroid (see Figure~\ref{categories}(a)). Expression values in the cluster with the higher mean centroid are assigned a Boolean value of \textbf{1}, while those in the cluster with the lower mean centroid are assigned \textbf{0}. Additionally, a region  can be defined at the midpoint between the two centroids to serve as a decision boundary. This approach has been implemented in an R package, as described in~\cite{blatte2019martin0,blatte2019martin1}.

\subsection{Density distribution estimation methods} 
The second method for quantifying gene expression data relies on estimating the probability density distributions of the measurement points. For each gene, a density distribution is estimated based on its expression measurements over time. Various approaches exist for identifying and approximating these distributions, as discussed in \cite{zhou2003binarization, beal2019personalization, jung2017refbool}. For instance, histogram plots of the measurement points provide a quick way to visualize the probability density. After the estimation, a bimodal distribution of each gene expression is expected to ensure effective binarization (see Figure~\ref{categories},(b)). This method involves modeling the measurements into two density distributions: one representing low expression values and the other representing high expression values. This approach is analogous to bipartition clustering and can yield similar binarization results.  The binarization is then based on the probability of being in the first distribution or in the second one, or according to a threshold defined using the two distribution parameters. At the step of binarization in \cite{tuna2009cross, tuna2010inference}, the authors focus on the approach given in \cite{zhou2003binarization}, which consists of modeling the measurements into two normal distribution densities with the estimation of their parameters. They find this method relatively more principled compared to the other approaches. In fact, the other methods use arbitrary thresholds; for example, in \cite{beal2019personalization}, if the probability of being in the first distribution is greater than (or equal to) 0.95, then the gene expression level is set to 0, and if the probability of being in the second distribution is greater than (or equal to) 0.95, then the gene expression level is set to 1, and all the others are not assigned (NA) otherwise. However, thresholds are based on the mean and the standard deviation of the two normal distributions estimated in \cite{zhou2003binarization}. For example, the threshold is located at the center of the two means when the two distributions have equal standard deviations. This method is also improved in \cite{tuna2010inference} for a possible trinarization by identifying from experimental data three Gaussian distribution  mixture models. Then, the three Gaussian distributions are sorted by means, and two thresholds are set. Each threshold is selected to be between the two adjacent Gaussian distributions and using the same formula as in \cite{zhou2003binarization}. 

\subsection{Fast dynamics localization methods}

The third approach for quantizing gene expression relies on detecting the fastest dynamics between two successive measurements in order to locate a threshold. Unlike the similarity-based strategies described above (e.g., clustering around a centroid or density distribution), this method focuses on identifying the strongest variation in the temporal evolution of expression values. Specifically, it detects the sharpest transitions between successive values, or ``jumps'', by measuring the distances between successive sorted data points. 

In \cite{shmulevich2002binary}, the authors proposed placing the threshold at the point of maximal separation between low and high values, which corresponds to the first large finite difference in the ordered data. Along similar lines, other methods approximate each gene expression profile by the step function that best fits the $n$ time points, assigning 0 to the lowest level and 1 to the highest. For example, \cite{hopfensitz2011multiscale} define a strong discontinuity as a high ratio between the jump size (Euclidean distance) and the approximation error of the step function with respect to the observed data (see Figure~\ref{categories}(c)). In this framework, a valid threshold requires the combination of a large jump size and a low fitting error. An implementation of this approach is available in R \cite{blatte2019martin0, blatte2019martin1}. 

A related strategy was proposed in \cite{sahoo2007extracting}, where thresholds are inferred by computationally fitting either one or two step functions. The algorithm systematically evaluates every possible step position between time points and selects the configuration that minimizes the squared error between the observed data and the candidate step function.

\subsection{Analysis of methods}

\paragraph{Dependence on measurement.}  
Binarization methods require a large number of measurements (ideally, time-series data with very small intervals) to define thresholds with high precision.  
However, experimental gene expression data are usually sparse and limited to a few snapshots \cite{edgar2002gene}.  
When only few  measurement times are available, thresholds are selected within wide uncertainty intervals.  
Including more measurements reduces this uncertainty, but when data are scarce, different methods may produce inconsistent results.  
This variability has been demonstrated in \cite{seguel2013semantics}, where different threshold reconstruction algorithms tested on two datasets led to significantly divergent outcomes.  

\paragraph{Non-bimodal distribution-based methods.}  
Approaches relying on density distributions also require many measurements to classify the expression values reliably.  
However,  the distribution of gene expression may not not always be bimodal.  
In such cases, thresholds defined with approximated methods, for example using the inter-quartile range (IQR) as in \cite{beal2019personalization}:  
values greater than or equal to the third quartile plus the IQR are set to 1,  
values less than or equal to the first quartile minus the IQR are set to 0,  
and intermediate values remain undefined.  
Currently, no robust binarization method exists for non-bimodal distributions, and applying such heuristics may yield unreliable results.  
The problem is further exacerbated by oscillatory or fluctuating gene expression patterns observed in many biological processes, including disease states \cite{szybinska2017p53, brady2010p53}.  

\paragraph{Dynamics-based methods.}  
Other methods focus on detecting fast dynamics in time-series data.  
Their applicability, however, is restricted to datasets with equally spaced measurement intervals. . In fact, how could we compare the dynamics speed of two discontinuities between two different time intervals of measurements? If the size intervals of measurements are not equal, then a strong discontinuity (or a distance) between two observation values could not be due to fast changes in the behavior but because the time interval between these two measurements or samplings is larger than the time interval of the other discontinuities. In fact, even if the time of each measurement is provided in addition, the ratio of the Euclidean distance between two observation values and the Euclidean distance between their time interval of measurements does not help too much, in particular, when there is a mixture of fast and slow dynamics between two observations and the time interval between the slow dynamics is too large compared to the time interval between the fast dynamics.

\paragraph{Continuous model reconstructions.}  
Alternative approaches, such as continuous model threshold reconstructions \cite{drulhe2008switching, porreca2008structural}, use perfect time-series datasets to detect switches between functional modes.  
Despite their sophistication, these methods still serve the basic quantization role: deciding whether a gene is active (on) or inactive (off).  
Here, the Boolean values have biological significance.
When modeling GRNs using Boolean formalism, the role of $0$ (the gene is off) and $1$ (the gene is on) is related to the ability of genes to regulate the expression of their targets. Thus, the Boolean values have functional roles: the gene is able to regulate the transcription of its targets or not. When a gene threshold value is defined using the experimental biological data, it should clearly separate between these two functional situations, and the gene expression level must correspond to the capacity to regulate the target. Thus, thresholds should not only reflect measurement values but also capture the functional role of genes in regulating transcription.

\paragraph{Conclusion.}  
In summary, the main challenge lies in applying reliable binarization methods to imperfect or incomplete datasets—where only a few measurements are available or where some gene expression values are missing altogether.  
Moreover, binarization should incorporate the functional role of genes in gene regulatory networks.  
In the following, we propose an original approach that addresses these limitations more effectively than existing methods.

\section{Bi4Back Algorithm}
\label{sec:Bi4Back}
In this section, we present an algorithm for gene expression data binarization using a novel method considered supervised, based on gene expression regulation, {\it i.e.}, it uses the gene functional roles.

This method can be applied to instantaneous gene expression data, even when only a single measurement is available, such as a steady-state snapshot. Conceptually, the approach infers binary gene activity states by combining thresholding, derived from the characterization of expression data, with functional binary completion. The completion process is guided by a traversal of the regulatory graph, where binary values are iteratively propagated from regulators to their targets according to Boolean regulatory rules. In this framework, assigning a value of 1 to a gene indicates that it is functionally active and contributes to the activation or inhibition of its downstream targets, whereas assigning a value of 0 denotes that the gene is inactive and exerts no regulatory effect.
 
 Thus, the attribution of a binary value using a threshold should comply with this functional distinction according to its state. Here, a gene is considered active when it is expressed and able to regulate the expression of its targets; otherwise, it is inactive. The switches are specifically dependent on a different threshold for each gene. Therefore, the threshold reconstruction is specific to each gene. To respect the functional role of each gene, the correction of the assigned binary values from the instantaneous data should be based on the analysis of the regulatory network. The outline of the algorithm is:

\begin{description}
\item[Initialization:] The role of this step is to define the binary values of some genes.
\item Forward consensus: The role is to complete, when possible, all the genes that have not yet been binarized by forwarding the Boolean values of regulators toward the target.
\item [Back propagating consensus:] The role is also to complete, when possible, the genes that have not yet been binarized by a back propagation of the Boolean values of the target toward its regulators.
\item [Harmonization:] The role is to assign a Boolean value to the regulators that have almost similar gene expression values.
\item[Inconsistency test:] The role is the correction of falsely assigned binary values from the instantaneous data.
\end{description}

The last four steps (from forwarding until the inconsistency test) are executed each time for each gene, and the iterations continue until a fixed point is reached, {\it i.e.}, until no further modifications of the binary values are possible. We illustrate in Figure~\ref{fig1} the main steps or the preview of our proposed algorithm :

\begin{figure}[htb!]
\begin{center}
\begin{tikzpicture} [scale=0.7]


\node at ( 3,0) [name=data, shape=ellipse,draw=black, rounded corners,
                           minimum width=4cm, minimum height=1cm,
                           text centered, scale=0.6]{\color{black}{Data \& Regulatory graph}} ; 
                           
\node  at ( 3,-2) [name=initialization, shape=rectangle,draw=black, rounded corners,
                           minimum width=4cm, minimum height=1cm,
                           text centered, scale=0.6]{\color{black}{Initialization}};         
 
 \node  at ( 3,-4) [name=Forward, shape=rectangle,draw=black, rounded corners,
                           minimum width=4cm, minimum height=1cm,
                           text centered, scale=0.6]{Forward Boolean propagation};     
                           
 \node  at ( 3,-6) [name=Backward, shape=rectangle,draw=black, rounded corners,
                           minimum width=4cm, minimum height=1cm,
                           text centered, scale=0.6]{Backward Boolean propagation};      
                      \node at ( 3,-8) [name=test, shape=diamond,draw=black, rounded corners,
                           minimum width=4cm, minimum height=1cm, text centered, scale=0.6] {Test};                                                                                                                           
                           
 \node at ( 3,-10) [name=output, shape=ellipse,draw=black, rounded corners,
                           minimum width=4cm, minimum height=1cm, text centered, scale=0.6] {Output: Binarized Profile};                                                                                                                                                     
                                                                    
\Edge[Direct, label={}, color= gray, bend=0, style={}](data)(initialization);       
\Edge[Direct, label={}, color= gray, bend=0, style={}](initialization)(Forward) ;    
\Edge[Direct, label={}, color= gray, bend=0, style={}](Forward)(Backward);  
\Edge[Direct, label={}, color= gray, bend=0, style={}](Backward)(test);    
\Edge[Direct, label={}, color= gray, bend=0, style={}](test)(output);  
\Edge[ label={}, color= gray, bend=0, style={}](test.0)(6.3,-8);      
\Edge[ label={}, color= gray, bend=0, style={}](6.3,-8)(6.3,-4);     
\Edge[Direct, label={}, color= gray, bend=0, style={}](6.3,-4)(5.5,-4);    
\node at (6.3,-6)  [right] {\rotatebox{90}{\tiny Different Boolean Profile}};   
\node at (3,-9)  [right] {\tiny Similar Boolean Profile };            
\end{tikzpicture}
\caption{The binarization process main steps.}
\label{fig1}
\end{center}
\end{figure}
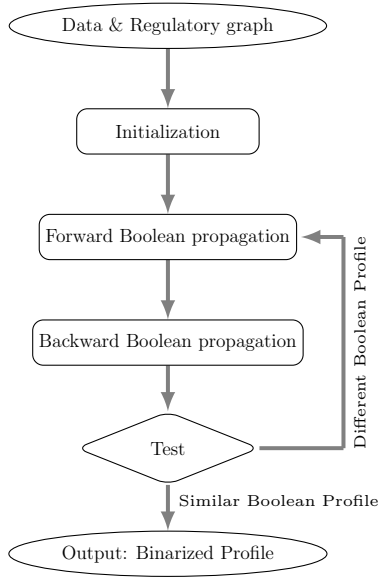

Methodologically, continuous gene expression data are normalized to the [0, 1] range using the Min–Max method \cite{patro2015normalization, yuvaraj2023lung, zhao2020quantile}, which we found most suitable as it scales all values within a common interval. After normalization, if certain genes represented in the regulatory graph lack corresponding expression measurements in the dataset (either missing or unassigned), they are assigned a neutral value of 0.5. This assignment facilitates the back-propagation of Boolean states from targets to regulators by enabling comparisons of expression levels across all regulators of a given gene.

In detail, the main steps of the binarization process are defined as follows.  

\subsection{Initialization} 
A binary profile for a subset of genes is generated from extreme expression values. The use of these extrema ensures that the corresponding genes can be reliably transformed into binary states. Accordingly, expression values very close to $0$ are assigned to binary level $0$, while genes exhibiting high normalized expression levels (close to $1$) are assigned to binary level $1$. Methodologically, the parameter $0 \leq \epsilon \leq 1$ is defined by the user, and the binary value $v_b$, derived from the continuous value $v$, follows the rules:
\begin{itemize}
	\item $v \leq \epsilon \implies v_b = 0;$
	\item $v \geq 1 - \epsilon \implies v_b = 1.$
\end{itemize}
In practice, $\epsilon$ should not exceed $0.05$.

\medskip
In addition, a complementary criterion is applied to genes associated with known biomarkers. In such cases, prior biological knowledge indicates that certain genes are necessarily active or inactive under specific experimental conditions, for instance, in a defined disease context such as a particular cancer type. For these experiments, gene activities are assigned according to the corresponding biomarker profile.. 

\subsection{Forward Consensus}
At this stage, the NA gene expression levels are completed by propagating the Boolean states of regulators to their respective targets within the regulatory graph, following the rules defined by the underlying Boolean network. Specifically, two core regulatory rules are applied to each node in the network:
\begin{itemize}
	\item If all inhibitors of a gene are expressed at level $1$ and all activators are at level $0$, the target gene is deterministically assigned a value of $0$.
	\item Conversely, if all activators are at level $1$ and all inhibitors are at level $0$, the target gene is deterministically assigned a value of $1$.
\end{itemize}

\begin{figure}
\begin{center}
\begin{tikzpicture}
\node  at ( 3,-1.25) [above,scale=0.7] {$s$};
\node  at ( 3,-1.5) [name=s,  shape=circle,draw, scale=0.5] 
{
$
\textcolor{red}{{\tiny Na}}
$
}; 
\node  at ( 0, 0.25) [above,scale=0.7] {\small {$Reg_{1}$}};
\node  at ( 0,0) [name=c, shape=circle,draw, scale=0.85] 
{$\tiny
\begin{aligned} 
\textcolor{red}{{\tiny 1}}
\end{aligned}
$
};
\node  at ( 0, -0.75) [above,scale=0.7] {\small $Reg_{2}$};
\node  at ( 0,-1)[name=d , shape=circle,draw, scale=0.85] { 
$
\tiny
\begin{aligned} 
 \textcolor{red}{\tiny 1}
\end{aligned}
$
};
\node  at ( 0, -1.75) [above,scale=0.7] {\small $Reg_{3}$};
\node  at ( 0,-2)[name=a, shape=circle,draw, scale=0.85] {
$
\tiny 
\begin{aligned} 
 \textcolor{red}{\tiny 0}
\end{aligned}
$
};
 \node  at ( 0, -2.75) [above,scale=0.7] {\small $Reg_{4}$};
\node  at (0,-3)[name=f,  shape=circle,draw, scale=0.85] {
$
\tiny
\begin{aligned} 
\textcolor{red}{\tiny 0}
\end{aligned}
$
};

\Edge[Direct, label={\textcolor{red}{$+$} }, fontcolor= ,bend=5, style={}](c)(s);


\Edge[Direct, label={\textcolor{red}{$+$}}, fontcolor= ,bend=5, style={}](d)(s);


\Edge[Direct, label={ \textcolor{red}{$-$} }, fontcolor={red}, bend=5, style={}](f)(s);

\Edge[Direct, label={ \textcolor{red}{$-$} }, fontcolor={red} ,bend=5, style={}](a)(s);

\end{tikzpicture}

\caption{An example where the target is not assigned and its regulators are defined. Here, $Reg_i$ represents the regulators of the target $s$.}
\label{Forward}
\end{center}
\end{figure}
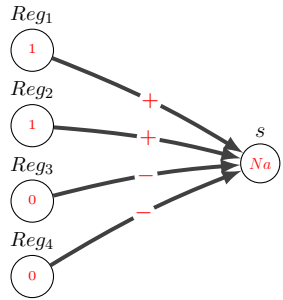

These rules are applied to any gene whose binary state has not yet been determined and whose regulators have defined values (see Figure~\ref{Forward}). The practical efficiency of this approach stems from the observation that most genes are regulated by a small number of regulators (typically $\leq 3$) of the same type. Consequently, scenarios such as both activators being set to $1$ occur relatively frequently, facilitating the propagation of expression states.

\subsection{Back Propagating Consensus}
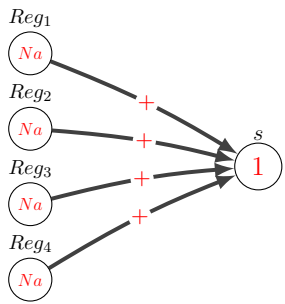
\begin{figure}
\begin{center}
\begin{tikzpicture}
\node  at ( 3,-1.25) [above,scale=0.7] {$s$};
\node  at ( 3,-1.5) [name=s, shape=circle,draw, scale=0.85] 
{
$
\textcolor{red}{{\tiny 1}}
$
}; 
\node  at ( 0, 0.25) [above,scale=0.7] {\small {$Reg_{1}$}};
\node  at ( 0,0) [name=c,  shape=circle,draw, scale=0.55] 
{
$
 \textcolor{red}{\tiny Na}
$
};

\node  at ( 0, -0.75) [above,scale=0.7] {\small $Reg_{2}$};
\node  at ( 0,-1)[name=d,  shape=circle,draw, scale=0.55] 
{
$
 \textcolor{red}{\tiny Na}
$
};
 \node  at ( 0, -1.75) [above,scale=0.7] {\small $Reg_{3}$};
\node  at ( 0,-2)[name=a,  shape=circle,draw, scale=0.55] {
$
 \textcolor{red}{\tiny Na}
$
};
 \node  at ( 0, -2.75) [above,scale=0.7] {\small $Reg_{4}$};
\node  at (0,-3)[name=f,  shape=circle,draw, scale=0.55] {
$
 \textcolor{red}{\tiny Na}
$
};

\Edge[Direct, label={\textcolor{red}{$+$} }, fontcolor= ,bend=5, style={}](c)(s);


\Edge[Direct, label={\textcolor{red}{$+$}}, fontcolor= ,bend=5, style={}](d)(s);


\Edge[Direct, label={ \textcolor{red}{$+$} }, fontcolor={red}, bend=5, style={}](f)(s);

\Edge[Direct, label={ \textcolor{red}{$+$} }, fontcolor={red} ,bend=5, style={}](a)(s);

\end{tikzpicture}
\caption{An example where the target is defined and all its regulators are not assigned (NA).}
\label{Backward}
\end{center}
\end{figure}

Unassigned (NA) gene expression levels are inferred by back-propagating the Boolean values from a defined target toward its regulators, under the condition that the target has a already defined state ($1$ or $0$) and all its consistent regulators are NA (see example in Figure~\ref{Backward}).
This back-propagation process relies on the notion of \emph{consistency}. A regulator is considered  consistent if its Boolean value alone can lead to the observed Boolean value of the target.
\begin{figure}
\begin{center}
\begin{tikzpicture}
\node  at ( 3,-1.25) [above,scale=0.7] {$s$};
\node  at ( 3,-1.5) [name=s,  shape=circle,draw, scale=0.85] 
{
$
\textcolor{black}{{\tiny 1}}
$
}; 
\node  at ( 0, 0.25) [above,scale=0.7] {\small \textcolor{teal} {$Reg_{1}$}};
\node  at ( 0,0) [name=c, shape=circle,draw, scale=0.57] 
{
$
 \textcolor{teal}{\tiny Na}
$
};

\node  at ( 0, -0.75) [above,scale=0.7] {\small \textcolor{teal} {$Reg_{2}$}};
\node  at ( 0,-1)[name=d,  shape=circle,draw, scale=0.85] 
{
$
 \textcolor{teal}{\tiny 0}
$
};
 \node  at ( 0, -1.75) [above,scale=0.7] {\small \textcolor{red} {$Reg_{3}$}};
\node  at ( 0,-2)[name=a, ball color = red, shape=circle,draw, scale=0.85] {
$
 \textcolor{black}{\tiny 0}
$
};
 \node  at ( 0, -2.75) [above,scale=0.7] {\small \textcolor{red} {$Reg_{4}$}};
\node  at (0,-3)[name=f, ball color = red, shape=circle,draw, scale=0.85] {
$
 \textcolor{black}{\tiny 1}
$
};

\Edge[Direct, label={\textcolor{blue}{$+$} }, fontcolor= , color=, bend=5, style={}](c)(s);


\Edge[Direct, label={\textcolor{blue}{$-$}}, fontcolor= , color=, bend=5, style={}](d)(s);


\Edge[Direct, label={ \textcolor{blue}{$-$} },  color=red, bend=5, style={}](f)(s);

\Edge[Direct, label={ \textcolor{blue}{$+$} },  color=red, bend=5, style={}](a)(s);
\end{tikzpicture}
\caption{An example where regulators 3 and 4 (shown in red) are not consistent with the target.}
\label{RegInc1}
\end{center}
\end{figure}
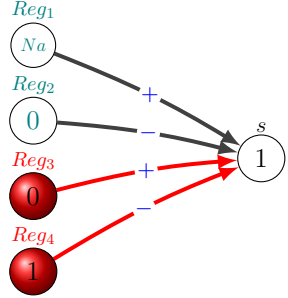
Consistency arises in one of the following situations: a target with value $1$ while an activator is $1$, or a target with value $0$ while an inhibitor is $1$. Conversely, consistency also occurs when a target is $0$ with an activator at $0$, or when a target is $1$ with an inhibitor at $0$. All other cases are considered inconsistent (see Figure~\ref{RegInc1}). The back-propagation procedure checks for the existence of a consistent rule that could explain the regulation, and potentially assigns binary values to the regulators to satisfy this consistency. 

Since the back-propagation step assigns a binary value to a single regulator, it is necessary to identify the most suitable candidate. This candidate is assumed to be the regulator that predominantly influences the target's expression.  

To address this issue, we introduce a set of techniques designed to determine, with reasonable confidence, which regulatory transition is actively acting on the target. Our approach relies on the gene expression values of the regulators to infer the active transition. Specifically, if the the target is set to $1$, we compare the regulator expression, assuming that the predominant regulator corresponds to the most strongly expressed activator or the most weakly expressed inhibitor.  

Formally, we define:

$$
\tau_i = 
\begin{cases} 
	B(t_j) (1 - \kappa_i) + \kappa_i (1 - B(t_j)), & \text{if } i \in \{1, \dots, a\}, \\[2mm]
	B(t_j) \kappa_i + (1 - \kappa_i) (1 - B(t_j)), & \text{if } i \in \{l - a, \dots, l\},
\end{cases}
$$

where $l$ denotes the total number of transitions regulating the target $t_j$, $B(t_j)$ is the Boolean value of the target, $a$ represents the number of activators, and $(l - a)$ the number of inhibitors. The expression value $\kappa_i$ corresponds to the measured expression level of each regulator. The dominant, or active, transition—whether activator or inhibitor—is defined as the one minimizing $\tau_i$ across all $i \in \{1, \dots, l\}$.

\subsection{Harmonization}

This step complements the back-propagation procedure. Considering all regulators of a given target, harmonization aims to assign Boolean states to regulators whose expression levels are closed to the regulators already assigned during back propagation. Specifically, for a target, if a regulator $i$ has been assigned an active Boolean state $s_i$, and there exists another regulator $j \in \{1, \dots, l\}$ such that $|\tau_i - \tau_j| < \delta$ (with $\delta$ a small threshold), then $s_j$ is set according to the regulatory interaction: $s_j = s_i$ if $i$ and $j$ act cooperatively, and $s_j = \neg s_i$ if their effects are non-cooperative.

\subsection{Inconsistency Test} 
\begin{figure}
\begin{center}
\begin{tikzpicture}
\node  at ( 3,-1.25) [above,scale=0.7] {\textcolor{red} {$s$}};
\node  at ( 3,-1.5) [name=s, ball color = red, shape=circle,draw, scale=0.85] 
{
$
\textcolor{black}{{\tiny 0}}
$
}; 
\node  at ( 0, 0.25) [above,scale=0.7] {\small \textcolor{red} {$Reg_{1}$}};
\node  at ( 0,0) [name=c, ball color = red, shape=circle,draw, scale=0.85] 
{
$
 \textcolor{black}{\tiny 1}
$
};

\node  at ( 0, -0.75) [above,scale=0.7] {\small \textcolor{red} {$Reg_{2}$}};
\node  at ( 0,-1)[name=d, ball color =red, shape=circle,draw, scale=0.85] 
{
$
 \textcolor{black}{\tiny 0}
$
};
 \node  at ( 0, -1.75) [above,scale=0.7] {\small \textcolor{red} {$Reg_{3}$}};
\node  at ( 0,-2)[name=a, ball color = red, shape=circle,draw, scale=0.85] {
$
 \textcolor{black}{\tiny 1}
$
};
 \node  at ( 0, -2.75) [above,scale=0.7] {\small \textcolor{red} {$Reg_{4}$}};
\node  at (0,-3)[name=f, ball color = red, shape=circle,draw, scale=0.85] {
$
 \textcolor{black}{\tiny 0}
$
};

\Edge[Direct, label={\textcolor{blue}{$+$} }, fontcolor= , color=red, bend=5, style={}](c)(s);


\Edge[Direct, label={\textcolor{blue}{$-$}}, fontcolor= , color=red, bend=5, style={}](d)(s);


\Edge[Direct, label={ \textcolor{blue}{$-$} },  color=red, bend=5, style={}](f)(s);

\Edge[Direct, label={ \textcolor{blue}{$+$} },  color=red, bend=5, style={}](a)(s);
\end{tikzpicture}
\caption{A confusion example because the target is inconsistent with all its regulators.}
\label{Confusion}
\end{center}
\end{figure}
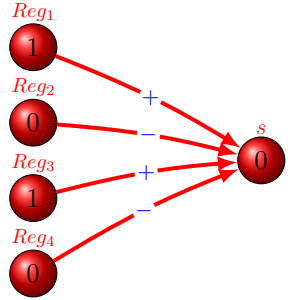

To respect the functional role of each gene, this step enables the correction of falsely assigned binary values from the instantaneous data. After the gene expression levels are assigned, an evaluation using the regulatory graph of the current Boolean value of the target should be performed to avoid confusions and the propagation of inconsistencies. In principle, at least one regulator should be consistent with the current Boolean value of the target. However, if the target is inconsistent with all its regulators, then a confusion exists according to the analysis of the regulatory network; see the example in Figure~\ref{Confusion}. 
In this case, we re-initialize some assigned genes to NA, which enables a new analysis for the Boolean values of these genes. Two possibilities could be considered for the re-initialization.

\begin{itemize}
\item The re-initialization of the Boolean value of the target because it is inconsistent with many regulators and, if it is inverted after the new analysis, then it becomes consistent with all its regulators. 

\item The re-initialization of the Boolean values of both the target and all its regulators.
\end{itemize}

We tested both possibilities, where the two choices provide almost similar results, but empirically, the second one is better in the presence of oscillations because it provides, in the end, more correct outcomes (binarized values). Therefore, we select a re-initialization of both the target and all its regulators in the rest of the paper.
So, the main steps of our proposed algorithm are summarized in Algorithm~\ref{algo1}. The algorithm stops when no new binary values are found. 

\begin{algorithm}
\caption{Forward-Backward Binarization Algorithm}\label{algo1}
\begin{algorithmic}[1]
\Require Normalized gene expression data \(D\), regulatory graph \(G\)
\State Initialize binary profile \(B\) with extremes and biomarkers
\While{a binary value is found}
    \State Forward propagate: Apply rules to undefined targets
    \State Backward propagate: Score regulators with Eq.~(1), assign dominant
    \State Harmonize: Assign similar regulators (\(|\tau_i - \tau_j| < \delta\))
    \State Test inconsistency: Re-initialize if target inconsistent with all regulators
\EndWhile
\Ensure Binarized profile
\end{algorithmic}
\end{algorithm}

\section{Experiments}

In this section, we benchmark our binarization algorithm using both real and artificial gene expression datasets. We begin by testing the algorithm on real gene expression data, specifically RNA-seq datasets, which are widely available from public databases such as the GDC Portal, TCGA, and GTEx. These databases provide accessible RNA gene expression data for various genes, tissues, or cancer types, enabling comprehensive evaluation of our method. Real experimental data are generated using technologies discussed in the introduction, ensuring relevance to practical biological applications. Additionally, we validate our algorithm using artificial datasets, which are derived from simulations of known biological models, including continuous ordinary differential equation (ODE) systems and discrete Boolean systems. Unlike real experimental data, artificial data are free from measurement noise, providing a controlled environment to assess the algorithm's performance. The artificial datasets are generated by simulating continuous ODE systems that represent either artificial gene regulatory networks (GRNs) or well-known Boolean biological networks. Each system is simulated over a time interval, and we extract snapshots (three experiments) after the system reaches a steady state. These snapshots are binarized using both the pre-set thresholds in the ODE system and our proposed binarization approach, allowing for a direct comparison of the results. Our method consistently produces binary profiles that align closely with those obtained using pre-set thresholds, demonstrating its robustness. To facilitate community access, the algorithm has been implemented in R and Mathematica, tested on real RNA-seq datasets, and validated using ODE simulations of both artificial and well-known Boolean GRNs. The source code is publicly available through Zenodo: \href{https://zenodo.org/records/10636447}{R} and \href{https://doi.org/10.5281/zenodo.11243896}{Mathematica} implementations.

\subsection{Testing the Algorithm}

\begin{figure}[htb!]
\begin{center}
\begin{tikzpicture}[
  node distance=2cm and 2cm,
  every node/.style={circle, draw=black, minimum size=1cm, inner sep=2pt, font=\small},
  activation/.style={-Stealth, thick, green!70!black, shorten >=2pt, shorten <=2pt},
  inhibition/.style={-|, thick, red!80!black, shorten >=2pt, shorten <=2pt}
]

\node[fill=blue!20!white] (EGFR) at (0,0) {EGFR};
\node[fill=blue!20!white] (ERK12) at (2,0) {ERK12};
\node[fill=blue!20!white] (PIK3CA) at (4,0) {PIK3CA};
\node[fill=blue!20!white] (AKT1) at (6,0) {AKT1};
\node[fill=blue!20!white] (GSK3) at (8,0) {GSK3};

\node[fill=blue!20!white] (MDM2) at (1,-2) {MDM2};
\node[fill=blue!20!white] (TP53) at (3,-2) {TP53};
\node[fill=blue!20!white] (PTEN) at (5,-2) {PTEN};
\node[fill=blue!20!white] (PARP1) at (7,-2) {PARP1};

\node[fill=blue!20!white] (BRCA1) at (2,-4) {BRCA1};
\node[fill=blue!20!white] (BCL2) at (4,-4) {BCL2};
\node[fill=blue!20!white] (BAX) at (6,-4) {BAX};
\node[fill=blue!20!white] (CCND1) at (8,-4) {CCND1};

\draw[activation] (EGFR) to (ERK12);
\draw[activation] (PIK3CA) to (AKT1);
\draw[activation] (TP53) to (PTEN);
\draw[activation] (ERK12) to (PARP1);
\draw[activation] (AKT1) to (BCL2);
\draw[activation] (TP53) to[bend left=30] (BAX);
\draw[activation] (PARP1) to (CCND1);
\draw[activation] (AKT1) to[bend right=30] (MDM2);
\draw[activation] (TP53) to[bend right=30] (MDM2);
\draw[inhibition] (EGFR) to[bend left=45] (BRCA1);
\draw[inhibition] (PTEN) to[bend right=45] (PIK3CA);
\draw[inhibition] (AKT1) to (GSK3);
\draw[inhibition] (MDM2) to[bend right=30] (TP53);
\draw[inhibition] (CCND1) to[bend left=45] (BRCA1);
\draw[inhibition] (BCL2) to (BAX);
\draw[inhibition] (GSK3) to[bend right=45] (CCND1);
\end{tikzpicture}
\caption{The interaction graph of breast cancer, focusing on central genes involved in tumor formation. Activation and inhibition are represented by green and red arrows, respectively.}
\label{RegulatoryGraph}
\end{center}
\end{figure}
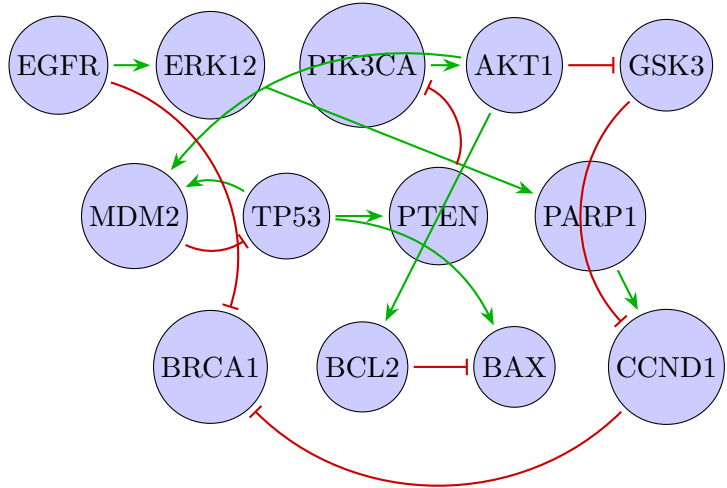

To evaluate the algorithm's performance on real-world data, we apply it to an RNA-seq gene expression dataset for ductal and lobular neoplasms, a subtype of breast cancer. The dataset, obtained from the GDC Portal (Metastatic Breast Cancer: MBC) \cite{GDC_case_MBCProject_0209}, contains transcriptomic profiles for genes critical to tumor formation. Based on literature and signaling pathway databases, we identified key regulatory interactions among these genes. For instance, \cite{kumaraswamy2015brca1} demonstrates that EGFR inhibits BRCA1 expression, while \cite{tsuchiya2014pi3} shows that PIK3CA activates AKT1. Using these interactions, a regulatory graph is constructed in \cite{biane2018causal, dehghan2022motif}, as depicted in Figure~\ref{RegulatoryGraph}.

The RNA-seq dataset includes six instantaneous experiments. We analyze the last experiment (final column), with gene expression values presented in Table~\ref{Table1}. These values are binarized using our proposed method, resulting in the binary profile shown in the table. While most genes are successfully binarized, some yield None values. This test was performed using Mathematica programming. Another example with a different regulatory graph was also tested using R programming on real-world data; we applied it to an RNA-seq gene expression dataset (available in \cite{Belgacem2024_R}). All genes were successfully binarized; for more details, see \cite{Belgacem2024_R}. To further assess the algorithm's reliability, we validate it using ODE simulations of artificial and well-known Boolean biological networks, as detailed in subsequent sections.

\begin{table}[htb!]
\centering
\small
\resizebox{\textwidth}{!}{
\begin{tabular}{|c|c|c|c|c|c|c|c|c|c|c|c|c|c|} 
\hline
Genes & EGFR & ERK12 & PIK3CA & AKT1 & GSK3 & MDM2 & TP53 & PTEN & PARP1 & BRCA1 & BCL2 & BAX & CCND1 \\ 
\hline 
Exp & 5.5972 & NaN & 7.073 & 27.5137 & 4.0499 & 17.098 & 2.324 & 30.655 & 10.7223 & 15.0989 & 0.2591 & 4.0974 & 9.147 \\ 
\hline 
Binary profile & False & False & False & False & True & None & None & False & None & None & False & False & False \\ 
\hline 
\end{tabular}
}
\caption{An RNA-seq experiment for ductal and lobular neoplasms and its binary profile using our binarization method.}
\label{Table1}
\end{table}

\subsection{The Algorithm Validation}

To rigorously validate our binarization algorithm, we evaluate its performance across a range of scenarios, starting with a simple artificial gene regulatory network exhibiting stable equilibria. This low-dimensional example allows us to illustrate how Boolean networks are modeled using ODE systems. We then generate artificial experimental data through ODE simulations and binarize these using our proposed approach, comparing the results to binary profiles derived from pre-set thresholds in the continuous models. We also investigate the algorithm's robustness against variations in model parameters. Subsequently, we extend the validation to high-dimensional, well-known Boolean biological networks \cite{biane2018causal, dehghan2022motif, Sahin2009, Traynard2016, Verlingue2016, Cohen2015}. Finally, we test the algorithm on an artificial GRN exhibiting oscillatory behavior to ensure its applicability to systems with dynamic fluctuations.

\subsubsection{An Artificial Example Exhibiting Stable States}

\begin{figure}[htb!]
\begin{center}
\begin{tikzpicture}[
  node distance=2cm,
  every node/.style={circle, draw=black, fill=white, minimum size=1cm, inner sep=2pt, font=\small},
  activation/.style={-Stealth, thick, green!70!black, shorten >=2pt, shorten <=2pt},
  inhibition/.style={-|, thick, red!80!black, shorten >=2pt, shorten <=2pt}
]

\node (g1) at (0,0) {$g_1$};
\node (g2) at (60:2.5cm) {$g_2$};
\node (g3) at (180:2.5cm) {$g_3$};
\node (g4) at (300:2.5cm) {$g_4$};
\node (g5) at (240:2.5cm) {$g_5$};

\draw[activation] (g1) to[bend left=15] (g2);
\draw[activation] (g2) to[bend left=15] (g4);
\draw[activation] (g3) to[bend left=15] (g1);
\draw[activation] (g3) to[bend left=15] (g5);
\draw[activation] (g4) to[bend left=15] (g1);
\draw[activation] (g5) to[bend left=15] (g1);

\draw[inhibition] (g2) to[bend right=15] (g3);
\draw[inhibition] (g4) to[bend left=15] (g5);

\end{tikzpicture}
\caption{An artificial gene regulatory network. Activation and inhibition are shown by green and red arrows, respectively.}
\label{fig2}
\end{center}
\end{figure}
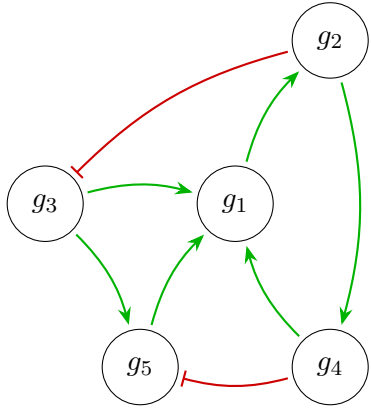

We begin with an artificial gene regulatory network, illustrated in Figure~\ref{fig2}, comprising five genes with defined interactions: gene \( g_1 \) activates \( g_2 \), \( g_2 \) activates \( g_4 \) and inhibits \( g_3 \), \( g_3 \) activates \( g_1 \) and \( g_5 \), \( g_4 \) activates \( g_1 \) and inhibits \( g_5 \), and \( g_5 \) activates \( g_1 \). These interactions are assumed to be cooperative and simultaneous, influencing each target gene in parallel. The Boolean network corresponding to this GRN is formalized in Equation~\eqref{Artmodel}, which defines the logical relationships governing gene activation and inhibition.

\begin{equation}\label{Artmodel}
\begin{array}{lcl}
g_1 & = & g_4 \wedge g_3 \wedge g_5 \\ [3mm]
g_2 & = & g_1 \\ [3mm]
g_3 & = & \neg g_2 \\ [3mm]
g_4 & = & g_2 \\ [3mm]
g_5 & = & g_3 \wedge \neg g_4 \\ [3mm]
\end{array}
\end{equation}

The Boolean model in Equation~\eqref{Artmodel} captures the logical dependencies among the genes, where \(\wedge\) denotes conjunction (AND), and \(\neg\) denotes negation (NOT). To model these dynamics continuously, we employ Hill functions, a standard approach for representing gene regulatory interactions in ODE systems \cite{farcot2019chaos}. The resulting ODE system, shown in Equation~\eqref{sys1}, describes the temporal evolution of gene product concentrations, incorporating activation and inhibition dynamics through increasing and decreasing Hill functions, respectively.

\begin{equation}\label{sys1}
\begin{array}{lcl}
\frac{dx_{1}}{dt} & = & \kappa_{1} h^+(x_3, \theta_3) h^+(x_4, \theta_4) h^+(x_5, \theta_5) - \gamma_{1} x_{1} \\ [3mm]
\frac{dx_{2}}{dt} & = & \kappa_{2} h^+(x_1, \theta_1) - \gamma_{2} x_{2} \\ [3mm]
\frac{dx_{3}}{dt} & = & \kappa_{3} h^-(x_2, \theta_2) - \gamma_{3} x_{3} \\ [3mm]
\frac{dx_{4}}{dt} & = & \kappa_{4} h^+(x_2, \theta_2) - \gamma_{4} x_{4} \\ [3mm]
\frac{dx_{5}}{dt} & = & \kappa_{5} h^-(x_4, \theta_4) h^+(x_3, \theta_3) - \gamma_{5} x_{5},
\end{array}
\end{equation}

In this system, \(x_i\) represents the concentration of the gene product for gene \(g_i\), \(h^+(x, \theta_i) = \frac{x^n}{x^n + \theta_i^n}\) is the increasing Hill function modeling activation, and \(h^-(x, \theta_i) = \frac{\theta_i^n}{x^n + \theta_i^n}\) is the decreasing Hill function modeling inhibition. The parameters \(\kappa_i\) and \(\gamma_i\) denote the expression and degradation rates, respectively, with degradation assumed proportional to concentration. The threshold \(\theta_i\) determines the transition point for gene activation or inhibition. This modeling approach, widely used in biological network analysis \cite{polynikis2009comparing, santillan2008use, belgacem2020control}, effectively bridges discrete Boolean logic with continuous dynamics.

\begin{figure}[htb!]
\begin{center}
\includegraphics[height=0.3\linewidth]{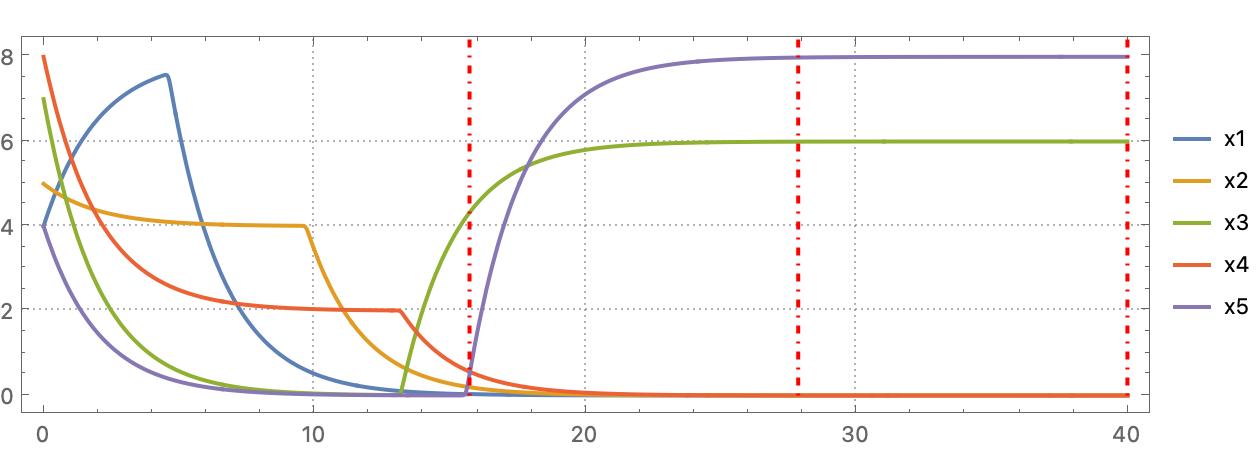}
\caption{The behavior of the system~\eqref{sys1} using Hill functions with parameters \(\kappa_1=4\), \(\kappa_2=2\), \(\kappa_3=3\), \(\kappa_4=1\), \(\kappa_5=4\), \(\gamma_1=\gamma_2=\gamma_3=\gamma_4=\gamma_5=0.5\), \(\theta_1=0.6\), \(\theta_2=0.7\), \(\theta_3=0.6\), \(\theta_4=0.6\), \(\theta_5=0.4\), and initial conditions \(x_{01}=4\), \(x_{02}=5\), \(x_{03}=7\), \(x_{04}=8\), \(x_{05}=4\).}
\label{fig3}
\end{center}
\end{figure}

A simulation of the system~\eqref{sys1} is shown in Figure~\ref{fig3}, illustrating convergence to a stable equilibrium. Three snapshots (\(Exp_1\), \(Exp_2\), \(Exp_3\)) are extracted near the steady state, marked by vertical red lines, and their values are reported in Table~\ref{T1}. These snapshots are binarized using both the pre-set thresholds (\(\theta_i\)) and our proposed method, with results compared to assess accuracy.

\begin{table}[t]
\centering
\scriptsize
\resizebox{\columnwidth}{!}{
\begin{tabular}{c c c c c c c}
\toprule
Genes & $g_1$ & $g_2$ & $g_3$ & $g_4$ & $g_5$ & d \\ 
\midrule
$Exp_1$ & 0.0329 & 0.197 & 4.31 & 0.563 & 0.5 & 0 \\ 
$Exp_2$ & 0 & 0.000532 & 6 & 0.0015 & 7.98 & 0 \\ 
$Exp_3$ & 0 & 0 & 6 & 0 & 8 & 0 \\ 
$\theta_i$ & 0.6 & 0.7 & 0.6 & 0.6 & 0.4 & \\ 
Binary profiles & False & False & True & False & True & \\ 
Binary profiles (our method) & False & False & True & False & True & \\ 
\bottomrule
\end{tabular}
}
\caption{Three artificial experiments of the system~\eqref{sys1} and their binary profiles using pre-set thresholds and our binarization method.}
\label{T1}
\end{table}

The binary profiles in Table~\ref{T1} demonstrate that our method perfectly matches the profiles obtained using pre-set thresholds, with a dissimilarity distance of \(d = \{0, 0, 0\}\). The dissimilarity distance is defined as the fraction of genes with differing binarized values between the two methods, where a gene with an NA value in our method is considered mismatched. This result confirms the accuracy of our binarization approach for stable systems. The Mathematica code for this example is available in \cite{Belgacem2024}.

\subsubsection{Study of Performance Against Parameter Variations}

To evaluate the robustness of our binarization method, we conducted 100,000 simulations of the system~\eqref{sys1} with randomly sampled parameters: \(\kappa_i \in [3, 100]\), \(\gamma_i \in [0.25, 2]\), and \(\theta_i = 1 + \delta\), where \(\delta \in [-0.5, 0.5]\). For each simulation, three snapshots were extracted near the steady state and binarized using our method and pre-set thresholds. Across all simulations, our method consistently produced identical binary profiles, with a dissimilarity distance of \(d = \{0, 0, 0\}\). The standard deviation of the sampled \(\kappa_i\) values was 27.98, indicating significant parameter variation, yet our method remained robust, highlighting its reliability across diverse conditions.

\subsubsection{Metastatic Breast Cancer Boolean Network Model}

We further validate our method using a Boolean network model of metastatic breast cancer \cite{biane2018causal}, with logical functions defined in Equation~\eqref{BreastBNF1} and illustrated in Figure~\ref{BreastBNF}. This model captures key regulatory interactions among genes involved in breast cancer progression, such as EGFR inhibiting BRCA1 and PIK3CA activating AKT1.

\begin{figure}[htb!]
\centering
\begin{equation}\label{BreastBNF1}
\begin{array}{lcl}
EGFR  & = & \neg BRCA1 \\ [3mm]
ERK12 & = & EGFR \\ [3mm]
PIK3CA & = & \neg PTEN \wedge EGFR \\ [3mm]
AKT1 & = & PIK3CA \\ [3mm]
GSK3B & = & \neg AKT1 \\ [3mm]
MDM2 & = & AKT1 \wedge TP53 \\ [3mm]
TP53 & = & \neg MDM2 \wedge (BRCA1 \vee \neg PARP1) \\ [3mm]
PTEN & = & TP53 \\ [3mm]
PARP1 & = & ERK12 \\ [3mm]
BRCA1 & = & \neg CCND1 \\ [3mm]
BCL2 & = & AKT1 \\ [3mm]
BAX & = & \neg BCL2 \wedge TP53 \\ [3mm]
CCND1 & = & (\neg GSK3B \wedge ERK12) \vee (\neg BRCA1 \wedge PARP1) \\ [3mm]
\end{array}
\end{equation}
\caption{The logical functions for the Boolean network of the metastatic breast cancer model \cite{biane2018causal}.}
\label{BreastBNF}
\end{figure}

The Boolean network has two stable steady states, as shown in Table~\ref{TableBrest}, representing distinct biological configurations of the system.

\begin{table}[htb!]
\centering
\small
\resizebox{\textwidth}{!}{
\begin{tabular}{|c|c|c|c|c|c|c|c|c|c|c|c|c|c|} 
\hline
Genes & EGFR & ERK12 & PIK3CA & AKT1 & GSK3 & MDM2 & TP53 & PTEN & PARP1 & BRCA1 & BCL2 & BAX & CCND1 \\ 
\hline 
$SST_1$ & True & True & True & True & False & False & False & False & True & False & True & False & True \\ 
\hline 
$SST_2$ & False & False & False & False & True & False & True & True & False & True & False & True & False \\ 
\hline 
\end{tabular}
}
\caption{Two Boolean steady states of the network in Figure~\ref{BreastBNF}.}
\label{TableBrest}
\end{table}

\begin{figure}[htb!]
\begin{center}
\includegraphics[height=0.3\linewidth]{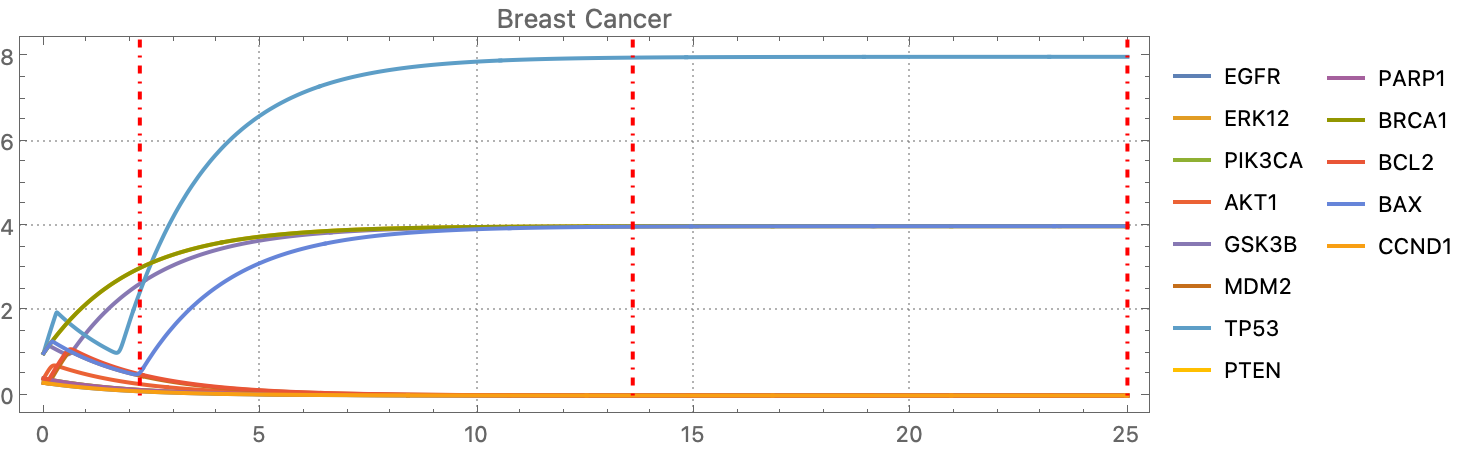}
\caption{Simulation of the ODE system for the Boolean network in Figure~\ref{BreastBNF} using Hill functions.}
\label{figBreast}
\end{center}
\end{figure}

Using Hill functions, we model this Boolean network as an ODE system, and a simulation converging to the first steady state (\(SST_1\)) is shown in Figure~\ref{figBreast}. Three snapshots are extracted near the steady state, marked by vertical red lines, with values reported in Table~\ref{Brest}. These are binarized using both pre-set thresholds and our method.

\begin{table}[htb!]
\centering
\small
\resizebox{\textwidth}{!}{
\begin{tabular}{|c|c|c|c|c|c|c|c|c|c|c|c|c|c|} 
\hline
Genes & EGFR & ERK12 & PIK3CA & AKT1 & GSK3 & MDM2 & TP53 & PTEN & PARP1 & BRCA1 & BCL2 & BAX & CCND1 \\ 
\hline 
$Exp_1$ & 3.06 & 3.09 & 2.22 & 2.28 & 0.452 & 0.246 & 0.115 & 0.246 & 3.98 & 0.131 & 3.55 & 0.0113 & 4.69 \\ 
\hline 
$Exp_2$ & 3.98 & 3.98 & 3.96 & 3.96 & 0.0101 & 0.00549 & 0.00257 & 0.00549 & 4.00 & 0.00294 & 3.99 & 0.000254 & 7.93 \\ 
\hline 
$Exp_3$ & 4.00 & 4.00 & 4.00 & 4.00 & 0.000223 & 0.000121 & 0 & 0.000121 & 4.00 & 0 & 4.00 & 0 & 8 \\ 
\hline 
$\theta_i$ & 0.503203 & 0.56938 & 0.362252 & 0.609277 & 0.456742 & 0.594652 & 0.462695 & 0.544637 & 0.445532 & 0.544715 & 0.508001 & 0.470601 & 0.472353 \\ 
\hline 
Binary profiles & True & True & True & True & False & False & False & False & True & False & True & False & True \\ 
\hline 
\end{tabular}
}
\caption{Three artificial experiments of the network in Figure~\ref{BreastBNF} and their binary profiles using pre-set thresholds.}
\label{Brest}
\end{table}

Our binarization method produces identical binary profiles to those obtained with pre-set thresholds, with a dissimilarity distance of \(d = \{0, 0, 0\}\), for more details, see the available code in \cite{Belgacem2024}. Notably, the real RNA-seq data for ductal and lobular neoplasms (Table~\ref{Table1}) align with the second steady state (\(SST_2\)), suggesting that our method accurately captures biologically relevant states.

\subsubsection{Boolean Biological Network Examples}

We further validate our approach using established Boolean models of biological regulatory networks, summarized in Table~\ref{tab:benchmark-networks}. These networks, derived from experimentally characterized systems, include up to 32 nodes and exhibit multiple steady states \cite{Cohen2015}.

\begin{table}[htb!]
\centering
\small
\begin{tabu} to \textwidth {X[3,c] X[5,l] X[1,c]}
\tabucline[1.5pt]{} 
Name of Model & Modeled Biological Process & Reference \\
\hline
Biane 2018 & Principal genes for metastatic breast cancer & \cite{biane2018causal} \\
Sahin 2009 & ERBB receptor-regulated G1/S transition network for anticancer drug analysis & \cite{Sahin2009} \\
Traynard 2016 & Mammalian cell cycle network & \cite{Traynard2016} \\
Verlingue 2016 & Signaling network controlling S-phase entry and geroconversion senescence & \cite{Verlingue2016} \\
Cohen 2015 & Regulatory network describing epithelial-to-mesenchymal transition & \cite{Cohen2015} \\
\tabucline[1.5pt]{}
\end{tabu}
\caption{effective Boolean biological networks.}
\label{tab:benchmark-networks}
\end{table}

For each model and steady state, we simulated the corresponding ODE system, extracted snapshots, and performed binarization. The resulting dissimilarity distances, reported in Table~\ref{Osci_0}, demonstrate excellent agreement with the expected profiles across most networks. Only a few genes are not binarized in the most complex, high-dimensional networks, highlighting the robustness and high accuracy of our method even in challenging scenarios.

\begin{table}[htb!]
\centering
\scriptsize
\resizebox{\columnwidth}{!}{
\begin{tabular}{|c|c|c|c|c|c|}
\hline
\textbf{Name of Model} & \textbf{Biane 2018} & \textbf{Sahin 2009} & \textbf{Traynard 2016} & \textbf{Verlingue 2016} & \textbf{Cohen 2015} \\
\hline
$SST_1$ & \{0,0,0\} & \{0,0,0\} & \{1/11,1/11,1/11\} & \{2/23,2/23,2/23\} & \{1/5,3/10,3/10\} \\
\hline
$SST_2$ & \{0,0,0\} & \{1/10,0,0\} & \{1/11,0,0\} & \{2/23,3/23,3/23\} & \{1/10,1/15,1/15\} \\
\hline
$SST_3$ & --- & \{1/4,0,0\} & --- & --- & \{1/10,1/15,1/15\} \\
\hline
\end{tabular}
}
\caption{Matching dissimilarity distances for each experiment, model, and stable state.}
\label{Osci_0}
\end{table}

\subsubsection{An Oscillating Artificial Example}

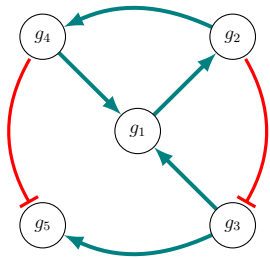
\begin{figure}[htb!]
\begin{center}
\begin{tikzpicture} 
\node at (0,0) [name=1, shape=circle, draw=black, rounded corners, minimum width=1cm, minimum height=1cm, text centered, scale=0.6]{$g_1$}; 
\node at (1.25,1.25) [name=2, shape=circle, draw=black, rounded corners, minimum width=1cm, minimum height=1cm, text centered, scale=0.6]{$g_2$};                               
\node at (1.25,-1.25) [name=3, shape=circle, draw=black, rounded corners, minimum width=1cm, minimum height=1cm, text centered, scale=0.6]{$g_3$};     
\node at (-1.25,1.25) [name=4, shape=circle, draw=black, rounded corners, minimum width=1cm, minimum height=1cm, text centered, scale=0.6]{$g_4$};                                 
\node at (-1.25,-1.25) [name=5, shape=circle, draw=black, rounded corners, minimum width=1cm, minimum height=1cm, text centered, scale=0.6]{$g_5$};  
\Edge[Direct, label={}, color=teal, bend=0, style={}](1)(2);  
\Edge[Direct, label={}, color=teal, bend=-25, style={}](2)(4);                         
\Edge[Direct, label={}, color=teal, bend=0, style={}](4)(1);                                                                                                                                
\path[-|, shorten <=1pt, shorten >=1pt, very thick, color=red] (4) edge [bend right] (5);
\path[-|, shorten <=1pt, shorten >=1pt, very thick, color=red] (2) edge [bend left] (3);
\Edge[Direct, label={}, color=teal, bend=25, style={}](3)(5);
\Edge[Direct, label={}, color=teal, bend=0, style={}](3)(1);
\end{tikzpicture}
\caption{An artificial gene regulatory network with oscillatory behavior.}
\label{OsciExample}
\end{center}
\end{figure}

Finally, we consider an artificial GRN exhibiting oscillatory behavior, shown in Figure~\ref{OsciExample}. The network structure is similar to Figure~\ref{fig2}, but with modified parameters to induce oscillations.

\begin{figure}[htb!]
\begin{center}
\includegraphics[height=0.4\linewidth]{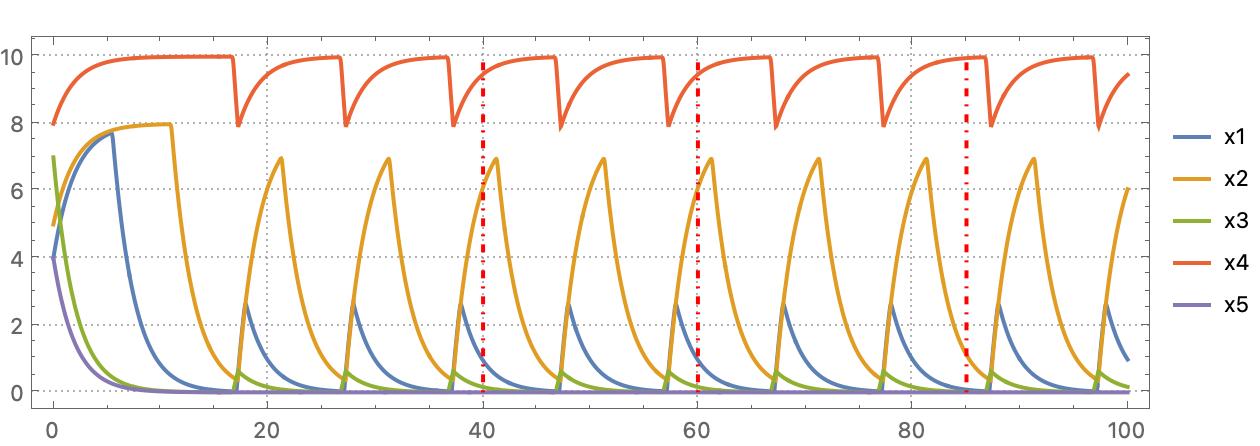}
\caption{The oscillatory behavior of the system with Hill functions (parameters \(\kappa_1=4\), \(\kappa_2=4\), \(\kappa_3=1.5\), \(\kappa_4=5\), \(\kappa_5=3\), \(\gamma_1=\gamma_2=\gamma_3=\gamma_4=\gamma_5=0.5\), \(\theta_1=0.5\), \(\theta_2=0.45\), \(\theta_3=0.45\), \(\theta_4=0.4\), \(\theta_5=0.6\), and initial conditions \(x_{01}=4\), \(x_{02}=5\), \(x_{03}=7\), \(x_{04}=8\), \(x_{05}=4\).}
\label{FigureOsci}
\end{center}
\end{figure}

The ODE system for this network, derived using Hill functions, exhibits oscillations under specific initial conditions and parameters. A simulation, shown in Figure~\ref{FigureOsci}, captures the oscillatory dynamics, with three snapshots extracted during a stable orbit (marked by red lines). These snapshots are binarized, and the results are reported in Table~\ref{Osci}.

\begin{table}[htb!]
\centering
\scriptsize
\resizebox{\columnwidth}{!}{
\begin{tabular}{lrrrrrr}
\toprule
\textbf{Genes} & \textbf{$g_1$} & \textbf{$g_2$} & \textbf{$g_3$} & \textbf{$g_4$} & \textbf{$g_5$} & \textbf{d} \\ 
\midrule
$Exp_1$ & 0.948409 & 6.11058 & 0.157608 & 9.47464 & 8.24$\times$10$^{-9}$ & 0 \\ 
$Exp_2$ & 0.958369 & 6.09073 & 0.159263 & 9.46912 & 3.74$\times$10$^{-13}$ & 0 \\ 
$Exp_3$ & 0.968433 & 6.07068 & 0.160936 & 9.46355 & 1.70$\times$10$^{-17}$ & 0 \\ 
$\theta_i$ & 0.5 & 0.45 & 0.45 & 0.4 & 0.6 & \\ 
Binary profiles & True & True & False & True & False & \\ 
\bottomrule
\end{tabular}
}
\caption{Three artificial experiments of the oscillatory network in Figure~\ref{OsciExample} and their binary profiles using pre-set thresholds.}
\label{Osci}
\end{table}

Our method accurately binarizes the oscillatory snapshots, matching the pre-set threshold profiles with a dissimilarity distance of \(d = \{0, 0, 0\}\),  for more details, see the available code in \cite{Belgacem2024}, confirming its effectiveness in dynamic systems.

 \section{Discussions}
The challenge of achieving confident binarization lies in selecting appropriate datasets. In our tests, we utilized transcriptomic data (RNA-seq) due to its widespread availability and the assumption that mRNA and protein expression levels are well-correlated, implying a Galois connection between mRNA and protein abundance. However, it is well-documented that mRNA and protein expression levels are not always correlated \cite{schwanhausser2011global, kristensen2013protein}. For instance, \cite{gillespie2020absolute} proposes a method to integrate temporal protein stoichiometry data with mRNA measurements to model mRNA dynamics as a function of regulator protein dynamics. Gene expression rates do not directly reflect mRNA or protein abundance, particularly when degradation occurs rapidly. The abundance of mRNA is determined by the balance between its synthesis and degradation rates. Moreover, mRNA expression levels may decrease even if an activator protein increases, or if the expression rate is high, due to post-transcriptional regulation. For example, mRNA may be targeted for degradation by newly expressed microRNAs, or a reduction in microRNA ``sponges'' may increase microRNA availability, leading to enhanced mRNA degradation \cite{liu2008control, wang2007microrna, mohr2015overview}. Thus, the expression rate alone does not fully capture the in vivo variation of mRNA levels.

To address this, we advocate for RNA-seq measurements conducted directly in vivo, as these reflect the cellular mRNA levels resulting from the balance of production and degradation, including effects such as dilution due to cell growth, division, or microRNA-mediated degradation \cite{romero2014rna}. In vivo RNA-seq measurements inherently account for these degradation processes, providing a more accurate representation of mRNA abundance at a specific time point.

Furthermore, in some cases, mRNA levels remain constant at equilibrium (where production equals degradation), and regulation occurs primarily at the translational or post-translational level \cite{schwanhausser2011global}. For instance, the abundance of proteins, particularly stress-responsive transcription factors, is often controlled at the translational level \cite{pakos2016integrated}. Many transcription factors are also regulated post-translationally through mechanisms such as phosphorylation, cleavage, sequestration, or proteolysis. Proteins may undergo enzymatic modifications post-synthesis, and their degradation can be triggered by processes like polyubiquitination, which tags transcription factors for degradation \cite{van2013stabilization}. Consequently, assuming that mRNA levels directly reflect protein levels is not always valid. For example, mRNA expression may increase while the corresponding protein level decreases due to variations in translation rates \cite{o2017protein, qu2011ribosome, makhoul2002distribution} or protein degradation triggered by post-translational modifications \cite{terman2013post}.

For confident binarization, we recommend using proteomic data measured in vivo, where the values reflect the degraded protein amounts, accounting for all degradation processes. If transcriptomic data are used, it is critical to ensure that measurements are taken in vivo and that regulation is primarily transcriptional, or that mRNA levels reliably indicate regulatory activity. When considering synthesis rates of mRNA or proteins in vivo, degradation rates must also be accounted for. The Harvard Database of Useful Biological Numbers provides valuable parameters for degradation and synthesis rates \cite{BioNumbers2009}. Additionally, studies such as \cite{chen2015genome} investigate RNA decay rates in \textit{Escherichia coli} using RNA-seq data, while \cite{neymotin2014determination} provides in vivo RNA degradation and synthesis rates. Similarly, \cite{kristensen2013protein, li2014quantifying} offer protein synthesis rates, and \cite{legewie2008recurrent} provides protein decay rates. In \cite{schwanhausser2011global}, mRNA and protein abundance for over 5,000 genes in mammalian cells were measured simultaneously, and a kinetic model was used to derive rate constants for mRNA and protein synthesis and degradation.

\section{Conclusion}

We present a novel binarization framework tailored for sparse gene expression datasets, including those limited to single-time-point measurements. Our method leverages Boolean regulatory logic to propagate binary states across the network, thereby resolving missing or incomplete data. Rigorous evaluation using real RNA-seq datasets and extensive validation through ordinary differential equation (ODE) simulations demonstrate perfect concordance with predefined thresholds.

Unlike conventional binarization techniques—which rely on multiple time-series measurements to establish robust thresholds and often fail with single steady-state snapshots—our regulation-based approach excels with instantaneous gene expression data. Comprehensive validation across diverse scenarios—encompassing synthetic gene regulatory networks, established Boolean biological models, and oscillatory systems—confirmed that our method consistently yields binary profiles with dissimilarity distances of zero, indicating exact alignment with theoretical expectations. Notably, across $100,000$ simulations with randomly perturbed parameters and substantial standard deviations, the algorithm maintained flawless accuracy.

A defining feature of our approach is its capacity to exploit network topology for inferring binary states in genes with missing expression values. By propagating information according to Boolean rules, the algorithm successfully binarizes genes that remain indeterminate under traditional threshold-based methods. Its robust performance extends from simple five-gene networks to complex models comprising up to 32 nodes and multiple steady states, sustaining high accuracy even under oscillatory dynamics.

Fundamentally, our method provides a critical preprocessing step for Boolean network inference, enabling the conversion of continuous gene expression data into the binary format required for downstream network analysis.

\bigskip
\textbf{Acknowledgments:} We thank Delphine Ropers (IBIS INRIA), Theodore J. Perkins (Ottawa Hospital Research Institute), and Catharine and Marc R. Roussel (Alberta RNA Research and Training Institute, University of Lethbridge) for discussions on gene regulation factors.

\clearpage 
\bibliographystyle{elsarticle-num}
\bibliography{mybibfile}

\end{document}